\documentclass[11pt]{article}
\usepackage{slashed}

\usepackage{amssymb}
\usepackage{graphics}
\usepackage{epsfig}
\usepackage{a4wide}
\usepackage[numbers,square,comma,sort&compress]{natbib}
\usepackage{color,soul}

\usepackage{xspace}
\usepackage{amsmath}
\usepackage{color,soul}

\begin{document}

\newcommand{\ggZHZH}{$gg \to Z_H Z_H~$}
\newcommand{\tabincell}[2]{\begin{tabular}{@{}#1@{}}#2\end{tabular}}
\newcommand\iden{\leavevmode\hbox{\small1\normalsize\kern-.33em1}}

\title{  Probing the littlest Higgs model with $T$ parity using di-Higgs events through $Z_H$-pair production at the LHC in NLO QCD }
\author{Chen Liang-Wen$^a$, Zhang Ren-You$^a$, Ma Wen-Gan$^a$, Li Wei-Hua$^a$, Duan Peng-Fei$^b$ and Guo Lei$^a$, \\
{\small  $^a$ Department of Modern Physics, University of Science and Technology of China, } \\
{\small  Hefei, Anhui 230026, People's Republic of China}  \\
{\small  $^b$ City College, Kunming University of Science and Technology, }  \\
{\small  Kunming£¬Yunnan 650051, People's Republic of China} }

\date{}
\maketitle \vskip 15mm
\begin{abstract}
We investigate the di-Higgs events through $Z_H$-pair production at the CERN Large Hadron Collider including the pure next-to-leading order (NLO) QCD correction and the $gg$-fusion contribution in the framework of the littlest Higgs model with $T$ parity. We employ the diagram subtraction scheme in the QCD NLO calculations to avoid double counting and keep the convergence of the perturbative QCD description for the $Z_H$-pair production. We investigate the dependence of the leading order and QCD corrected integrated cross sections on the renormalization/factorization scale, and find that the total QCD corrections slightly reduce the scale uncertainty in the plotted range. By considering the subsequent decays of the intermediately produced $Z_H$ bosons and adopting the exclusive four-$b$-jet event selection criterion, the QCD correction provides considerable enhancement of the kinematic distributions for final decay products. We find that it is possible to select the signature of the $Z_H$-pair production from possible standard model background by taking proper kinematic cuts.
\end{abstract}

 \vfill
\eject \vfill \eject

\baselineskip=0.32in

\renewcommand{\theequation}{\arabic{section}.\arabic{equation}}
\renewcommand{\thesection}{\Roman{section}}
\newcommand{\nb}{\nonumber}

\newcommand{\Dir}{\kern -6.4pt\Big{/}}
\newcommand{\Dirin}{\kern -10.4pt\Big{/}\kern 4.4pt}
\newcommand{\DDir}{\kern -7.6pt\Big{/}}
\newcommand{\DGir}{\kern -6.0pt\Big{/}}

\makeatletter
\@addtoreset{equation}{section}
\makeatother

\vskip 5mm
\section{INTRODUCTION}
\par
In 2012 both the ATLAS and CMS collaborations at the Large Hadron Collider (LHC) announced the observation of a new boson with mass of about $125 ~{\rm GeV}$. The present analyses indicate that this particle is compatible with the standard model (SM) Higgs boson \cite{Aad:2012tfa,Chatrchyan:2012ufa}. This discovery is a tremendous achievement in the history of particle physics. However, it is just the first step in understanding the electroweak symmetry breaking. Further precise investigations of the SM-like Higgs boson are in great demand and the existence of the new physics beyond the SM at the TeV energy scale is still an open issue.

\par
The littlest Higgs model (LHM) is an elegant realization of the little Higgs mechanism, which is proposed to ameliorate the fine-tuning problem \cite{ArkaniHamed:2002qy,ArkaniHamed:2001nc,ArkaniHamed:2002qx}. As the LHM suffers stringently from the electroweak precision constraints \cite{Schmaltz:2005ky,Perelstein:2005ka,
Hewett:2002px,Csaki:2002qg,Chen:2003fm,Han:2004az,Kilian:2003xt,
Low:2004xc,Hubisz:2004ft,Hubisz:2005tx,Cheng:2003ju}, a discrete symmetry, denoted as `$T$ parity', is implemented to enlarge the symmetry of the model. Under $T$ parity transformation, the SM particles are $T$ even and all the new heavy particles predicted in this model are $T$ odd except $T_+$. Therefore, the mixture of the SM gauge bosons with the new heavy gauge bosons is prohibited by this $T$ parity and the vacuum expectation value (VEV) of the weak-triplet scalar field vanishes. Consequently, the significant experimental constraints associated with the LHM are alleviated.

\par
The dark matter, which remains one of the most puzzling enigmas of the current fundamental physics, has gained significant attention. According to the current research, the dark matter should be cold and weakly interacting, typically detected as missing-energy signals at particle colliders. The littlest Higgs model with $T$ parity (LHT) predicts a neutral and colorless $T$-odd particle $A_H$ that can be a good candidate for dark matter \cite{Mitsou:2014wta,Birkedal:2006fz,Asano:2006nr}.

\par
The LHT phenomenology has been extensively studied \cite{lhtph,Belyaev:2006jh,Blanke:2006eb}, and the constraints from the LHC data have been examined and updated in Refs.{\cite{Han:2013ic,Reuter:2013iya,Reuter:2012sd}}. The heavy gauge boson pair productions at the leading order (LO) have been sketched in Ref.{\cite{Belyaev:2006jh}}. Recently the detailed higher order QCD corrections for $W_H$-pair and $W_HZ_H$ productions have been presented in Refs.{\cite{SongMing:2012gb,Wen:2013xb}.

\par
The di-Higgs boson production is often discussed as a probe of new physics, which plays a key role in probing the Higgs self-coupling and the existence of heavier states coupled to the Higgs boson. The goal of this paper is to examine QCD quantum effects on the $Z_H$-pair production with the subsequent $Z_H \to A_Hh \to A_Hb\bar{b}$ decay. Then it offers a di-Higgs boson production mechanism as illustrated in Fig.\ref{fig1}. Furthermore, we present detailed kinematic distributions for final products. The matrix elements of the process are calculated by adopting the developed \texttt{FeynArts}/\texttt{FormCalc}/\texttt{LoopTools} packages \cite{Hahn:1998yk,Hahn:2000kx,Hahn:2004rf} with the `t Hooft-Feynman gauge employed, and the numerical results at the LO are in agreement with those by using the CalcHep \cite{Pukhov:2004ca,Belyaev:2006jh}. The paper is organized as follows: The overview of the LHT theory is given in Sec.II. In Sec.III, we describe the calculation strategy for the $Z_H$-pair production at a proton-proton collider. The integrated and differential cross sections are provided and discussed in Sec.IV. Finally a summary is given.
\begin{figure*}
\begin{center}
\includegraphics[scale=0.6]{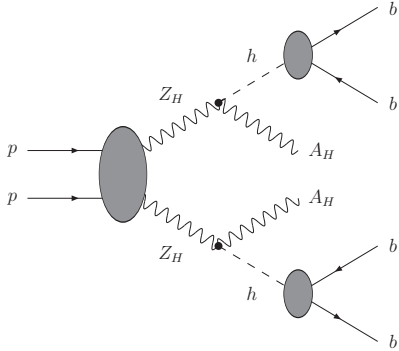}
\caption{ The production structure for the $b\bar{b}b\bar{b}+\slashed{E}_{T}$  signature via $Z_H$-pair production at the hadron collider. } \label{fig1}
\end{center}
\end{figure*}

\vskip 5mm
\section{OVERVIEW OF THE LHT}
\par
The details of the LHT theory can be found in Refs.\cite{Hubisz:2004ft,Hubisz:2005tx,Belyaev:2006jh,Blanke:2006eb}. The LHT is based on an $SU(5)/SO(5)$ nonlinear sigma model with an additional discrete symmetry, $T$ parity. The $SU(5) \rightarrow SO(5)$ global symmetry breaking leads to 14 massless Nambu-Goldstone bosons described by the ``pion'' matrix as
\begin{eqnarray}
\label{pion matrix}
\Pi =
\left(
\begin{array}{ccccc}
-\frac{\omega^0}{2} - \frac{\eta}{\sqrt{20}} &
-\frac{\omega^+}{\sqrt{2}} &
-i\frac{\pi^+}{\sqrt{2}} &
-i\phi^{++} &
-i\frac{\phi^+}{\sqrt{2}} \\
-\frac{\omega^-}{\sqrt{2}} &
\frac{\omega^0}{2} - \frac{\eta}{\sqrt{20}} &
\frac{v+h+i\pi^0}{2} &
-i\frac{\phi^+}{\sqrt{2}} &
\frac{-i\phi^0+\phi^P}{\sqrt{2}} \\
i\frac{\pi^-}{\sqrt{2}} &
\frac{v+h-i\pi^0}{2} &
\sqrt{4/5}\eta &
-i\frac{\pi^+}{\sqrt{2}} &
\frac{v+h+i\pi^0}{2} \\
i\phi^{--} &
i\frac{\phi^-}{\sqrt{2}} &
i\frac{\pi^-}{\sqrt{2}} &
-\frac{\omega^0}{2}-\frac{\eta}{\sqrt{20}} &
-\frac{\omega^-}{\sqrt{2}} \\
i\frac{\phi^-}{\sqrt{2}} &
\frac{i\phi^0+\phi^P}{\sqrt{2}} &
\frac{v+h-i\pi^0}{2} &
-\frac{\omega^+}{\sqrt{2}} &
\frac{\omega^0}{2} - \frac{\eta}{\sqrt{20}}
\end{array}\right).
\end{eqnarray}
This symmetry breaking takes place at the scale $f \sim {\cal O}({\rm TeV})$ and originates from the VEV of
the nonlinear sigma model field $\Sigma$, where the $SU(5)$ symmetric tensor field $\Sigma$ is described by
\begin{eqnarray}
\Sigma = e^{i \Pi/f} \Sigma_0 e^{i \Pi^{T}/f} = e^{2 i \Pi/f} \Sigma_0
\end{eqnarray}
with
\begin{eqnarray}
\Sigma_0 = \langle \Sigma \rangle =
\left(
\begin{array}{ccc}
&& 1_{2 \times 2} \\
& 1 & \\
1_{2 \times 2} &&
\end{array}
\right).
\end{eqnarray}

\subsection{T-odd heavy gauge bosons and scalars}
\par
An $\left[SU(2) \otimes U(1) \right]_1 \otimes\left[SU(2) \otimes U(1) \right]_2$ subgroup of the $SU(5)$ global symmetry is gauged, and
the gauge fields $W_{i \mu}^a$ and $B_{i \mu}$ $(a = 1, 2, 3,~ i = 1, 2)$ are introduced correspondingly. The kinetic terms for the gauge
and scalar fields can be written as
\begin{eqnarray}
 {\cal L}_{{\rm gauge+scalar}} =
 \sum_{i=1}^2
 \left[
 -\frac{1}{2} {\rm Tr} \Big( W_{i\mu\nu}W_i^{\mu\nu} \Big)
 -
 \frac{1}{4} B_{i\mu\nu}B_i^{\mu\nu}
 \right]
 +
 \frac{f^2}{8} {\rm Tr}
 \left[
 \Big( D_{\mu}\Sigma \Big)^{\dag} \Big( D^{\mu} \Sigma \Big)
 \right].
\end{eqnarray}
The covariant derivative $D_{\mu} \Sigma$ and the gauge field strength tensors $W_{i\mu\nu}$, $B_{i\mu\nu}$ $(i = 1, 2)$ are defined as
\begin{eqnarray}
&& D_{\mu} \Sigma
=
\partial_{\mu} \Sigma - i \sqrt{2} \sum_{i=1}^2
\left[
g \Big(
W_{i \mu} \Sigma + \Sigma W_{i \mu}^{T} \Big) + g^{\prime} B_{i \mu}
\left( Y_i \Sigma + \Sigma Y_i \right)
\right], \nonumber \\
&& W_{i \mu \nu} = \partial_{\mu} W_{i \nu} - \partial_{\nu} W_{i \mu} - i \sqrt{2} g \left[ W_{i \mu},~ W_{i \nu} \right], \nonumber \\
&& B_{i \mu \nu} = \partial_{\mu} B_{i \nu} - \partial_{\nu} B_{i \mu},
\end{eqnarray}
where $W_{i \mu} = W_{i \mu}^a Q_i^a$, and $Q_i^a$ and $Y_i$ $(a = 1, 2, 3,~ i = 1, 2)$ are the generators of the
$\left[SU(2) \otimes U(1) \right]_1 \otimes\left[SU(2) \otimes U(1) \right]_2$ gauge group,
\begin{eqnarray}
&&Q_1^a =
\frac{1}{2} \left(
\begin{array}{ccc}
\tau^a~ &~ &~ \\
&& \\
&&
\end{array}
\right)_{5 \times 5}~,
~~~~
Y_1 = \frac{1}{10} {\rm diag}\left(3, 3, -2, -2, -2\right),
\nonumber \\
&&Q_2^a =
\frac{1}{2} \left(
\begin{array}{ccc}
&& \\
&& \\
&& -\tau^{a \ast}
\end{array}
\right)_{5 \times 5}~,
~~~~
Y_2 = \frac{1}{10} {\rm diag}\left(2, 2, 2, -3, -3\right).
\end{eqnarray}
Under $T$ parity, the gauge and scalar fields transform as
\begin{eqnarray}
B_1 \longleftrightarrow B_2,~~~~~~ W_1^a \longleftrightarrow W_2^a,~~~~~~ \Pi \longrightarrow -\Omega \Pi \Omega,
\end{eqnarray}
where $\Omega = {\rm diag}\left(1,1,-1,1,1\right)$, while the Lagrangian ${\cal L}_{{\rm gauge+scalar}}$ is invariant.

\par
The VEV $\Sigma_0$ breaks the gauge symmetry $\left[SU(2) \otimes U(1) \right]_1 \otimes\left[SU(2) \otimes U(1) \right]_2$
to its diagonal $T$-even $SU(2) \otimes U(1)$ subgroup, with the generators
\begin{eqnarray}
Q^a = Q_1^a + Q_2^a,~~~~~~~ Y = Y_1 + Y_2.
\end{eqnarray}
This subgroup is identified with the SM electroweak gauge group, and usually denoted as $SU(2)_L \otimes U(1)_Y$. After the electroweak symmetry breaking $SU(2)_L \otimes U(1)_Y \rightarrow U(1)_{{\rm em}}$ via the SM Higgs mechanism, the mass eigenstates of the gauge sector are given by
\begin{eqnarray}
&& \left(
\begin{array}{c}
A_H \\
Z_H
\end{array}
\right)
=
\left(
\begin{array}{cc}
\cos\theta_H & -\sin\theta_H \\
\sin\theta_H & \cos\theta_H
\end{array}
\right)
\left(
\begin{array}{cccc}
\frac{1}{\sqrt{2}} & -\frac{1}{\sqrt{2}} & 0 & 0 \\
0 & 0 & \frac{1}{\sqrt{2}} & -\frac{1}{\sqrt{2}}
\end{array}
\right)
\left(
\begin{array}{c}
B_1 \\
B_2 \\
W_1^3 \\
W_2^3
\end{array}
\right), \nonumber \\
&& ~~~~ W_H^{\pm}
=
\frac{\left(W_1^1 - W_2^1\right) \mp i \left(W_1^2 - W_2^2\right)}{2}, \nonumber \\
&& \left(
\begin{array}{c}
A_L \\
Z_L
\end{array}
\right)
=
\left(
\begin{array}{cc}
\cos\theta_W & \sin\theta_W \\
-\sin\theta_W & \cos\theta_W
\end{array}
\right)
\left(
\begin{array}{cccc}
\frac{1}{\sqrt{2}} & \frac{1}{\sqrt{2}} & 0 & 0 \\
0 & 0 & \frac{1}{\sqrt{2}} & \frac{1}{\sqrt{2}}
\end{array}
\right)
\left(
\begin{array}{c}
B_1 \\
B_2 \\
W_1^3 \\
W_2^3
\end{array}
\right), \nonumber \\
&& ~~~~ W_L^{\pm} =
\frac{\left(W_1^1 + W_2^1\right) \mp i \left(W_1^2 + W_2^2\right)}{2},
\end{eqnarray}
where $A_H$, $Z_H$, and $W_H^{\pm}$ are $T$-odd heavy gauge bosons, while the $T$-even light ones, $A_L$, $Z_L$, and $W_L^{\pm}$, are identified with the SM photon and $Z$-, $W$-bosons.

\par
Among the 14 Nambu-Goldstone bosons given by the $\Pi$ matrix, $\eta$, $\omega^0$, and $\omega^{\pm}$ are the Goldstone bosons associated with the spontaneous gauge symmetry breaking $\left[SU(2) \otimes U(1) \right]_1 \otimes\left[SU(2) \otimes U(1) \right]_2 \rightarrow SU(2)_L \otimes U(1)_Y$, and are eaten by the heavy gauge bosons $A_H$, $Z_H$, and $W^{\pm}_H$, respectively. The remaining ten are classified into (1) a $T$-even SM Higgs doublet $H \sim \left(\pi^{+}, h+v, \pi^0\right)$ and (2)  a $T$-odd scalar triplet $\Phi \sim \left(\phi^{++}, \phi^{+}, \phi^0, \phi^P\right)$, where $h$ is the SM Higgs boson, $v$ the Higgs VEV, and $\pi^{0, \pm}$ are the Goldstone bosons eaten by the SM gauge bosons.

\par
The SM photon $A_L$ is massless due to the $U(1)_{{\rm em}}$ gauge invariance. The $T$-parity conservation ensures that the custodial relation $ m_{W_L} = m_{Z_L} \cos\theta_W$ is exactly satisfied at the tree level. At the ${\cal O}(v^2/f^2)$, the masses of gauge bosons are given by \footnote{From the expression for the $W$-boson mass, we obtain $v_{SM} = v \left( 1 - \frac{1}{12}\frac{v^2}{f^2} \right)$ at the ${\cal O}(v^2/f^2)$ \cite{Reuter:2012sd}.}
\begin{eqnarray}
m_{A_H} = \frac{1}{\sqrt{5}} g^{\prime} f \left( 1 - \frac{5}{8}\frac{v^2}{f^2} \right),~~~~
m_{Z_H} = m_{W_H} = g f \left( 1 - \frac{1}{8}\frac{v^2}{f^2} \right),~~~
m_{W_L} = \frac{1}{2} g v \left( 1 - \frac{1}{12}\frac{v^2}{f^2} \right),~~~
\end{eqnarray}
and the mixing angle $\theta_H$ has the form 
\begin{eqnarray}
\sin\theta_H = \frac{5 g g^{\prime}}{4 (5 g^2 - g^{\prime 2})} \frac{v^2}{f^2}.
\end{eqnarray}
For the scalar triplet $\Phi$, all the components are degenerate at the ${\cal O}(v^2/f^2)$ with the mass of
\begin{eqnarray}
m_{\Phi} = \sqrt{2} m_h \frac{f}{v},
\end{eqnarray}
where $m_h$ is the mass of the SM Higgs scalar.

\subsection{T-odd mirror fermions}
\par
To implement $T$ parity in the fermion sector, we introduce the following two incomplete left-hand $SU(5)$ multiplets and a right-hand $SO(5)$ multiplet:
\begin{eqnarray}
 \Psi_1
 =
 \left(
 \begin{array}{c}
 \psi_1 \\ 0 \\ 0
 \end{array}
 \right),~~~~
 \Psi_2
 =
 \left(
 \begin{array}{c}
 0 \\ 0 \\ \psi_2
 \end{array}
 \right),~~~~
 \Psi_{HR}
 =
 \left(
 \begin{array}{c}
 \tilde{\psi}_{HR} \\ \chi_{HR} \\ \psi_{HR}
 \end{array}
 \right),
\end{eqnarray}
with
\begin{eqnarray}
\psi_A = -\tau^2 q_A = -\tau^2 (u_A,~ d_A)^T,~~~~(A = 1, 2, HR),
\end{eqnarray}
for each fermion flavor. The transformations for these fields under the global $SU(5)$ are 
\begin{eqnarray}
\Psi_1 \longrightarrow V^{*} \Psi_1,~~~~~
\Psi_2 \longrightarrow V \Psi_2,~~~~~
\Psi_{HR} \longrightarrow U \Psi_{HR},
\end{eqnarray}
where $V \in SU(5)$ and $U$ is an $SO(5)$ transformation in a nonlinear representation of $SU(5)$. It tells us that $q_1$, $q_2$, and $q_{HR}$ are all $SU(2)_L$ doublets. Under $T$ parity, $\Psi_1$, $\Psi_2$, and $\Psi_{HR}$ transform as
\begin{eqnarray}
\Psi_1 \longrightarrow -\Sigma_0 \Psi_2,~~~~~
\Psi_2 \longrightarrow -\Sigma_0 \Psi_1,~~~~~
\Psi_{HR} \longrightarrow -\Psi_{HR}.
\end{eqnarray}
Thus, the $T$ parity eigenstates of the $SU(2)_L$ fermion doublets $q_A~ (A = 1, 2, HR)$ are given by
\begin{eqnarray}
&& q_{SM} = \frac{q_1 - q_2}{\sqrt{2}},~~~~~~~~~~~~~~~~~~~~~~~(T-{\rm even}), \nonumber \\
&& q_{HL} = \frac{q_1 + q_2}{\sqrt{2}},~~~~~~~ q_{HR},~~~~~~~~~~(T-{\rm odd}).
\end{eqnarray}
$q_{SM}$ is the left-hand $SU(2)_L$ SM fermion doublet, while $q_{HL}$ the left-hand $SU(2)_L$ mirror fermion doublet. The right-hand $SU(2)_L$ mirror fermion doublet is given by $q_{HR}$.

\par
The $T$-odd mirror fermions acquire masses via the following $SU(5)$ and $T$ parity invariant Lagrangian:
\begin{eqnarray}
 {\cal L}_{\rm mirror}
 =
 -\sum_{i,j=1}^3
\kappa_{ij} f
 \Big(
 \bar{\Psi}_2^i \xi +
 \bar{\Psi}_1^i \Sigma_0 \Omega \xi^{\dag} \Omega
 \Big)
 \Psi_{HR}^j ~+ ~{\rm h.c.},
\end{eqnarray}
where $i,j=1,...,3$ are flavor indices, and $\xi = e^{i \Pi/f}$, transforming under $SU(5)$ as
\begin{eqnarray}
\xi \rightarrow V \xi U^{\dag} = U \xi \Sigma_0 V^{T} \Sigma_0.
\end{eqnarray}
By assuming a diagonal and flavor independent coupling matrix, i.e., $\kappa_{ij} = \kappa \delta_{ij}$, the masses of the $T$-odd up- and down-type mirror fermions at the ${\cal O}(v^2/f^2)$ are given by
\begin{eqnarray}
m_{u_{i-}} = \sqrt{2}\kappa f\left( 1 - \frac{1}{8} \frac{v^2}{f^2} \right),~~~~~~
m_{d_{i-}} = \sqrt{2}\kappa f.
\end{eqnarray}
For the $T$-odd mirror quarks, $u_{i-} = u_-, c_-, t_-$ and $d_{i-} = d_-, s_-, b_-$ with $i$ running from 1 to 3.

\subsection{Top-quark partners}
\par
In order to cancel the large quadratic divergence to the Higgs boson mass induced by the top quark, a $T$-even top-quark partner $T_+$ is introduced. The implementation of $T$ parity then requires also a $T$-odd partner $T_-$. To properly describe the top sector particle content, the following multiplets are introduced:

\par
(1) Two left-hand $SU(5)$ multiplets $Q_1$ and $Q_2$,
\begin{eqnarray}
 Q_1 = \left( \begin{array}{c} \psi_1 \\ U_{L1} \\ 0
 \end{array}
 \right),~~~~~~~
 Q_2 =
 \left( \begin{array}{c} 0 \\ U_{L2} \\ \psi_2
 \end{array} \right).
\end{eqnarray}
\par
(2) Three right-hand $SU(2)_L$ singlets $U_{R1}$, $U_{R2}$, and $u_R$. \\
$Q_1$ and $Q_2$ obey the same transformation laws under $T$ parity and $SU(5)$ as do $\Psi_1$ and $\Psi_2$, respectively, and $U_{R1}$, $U_{R2}$, and $u_R$ transform under $T$ parity as
\begin{eqnarray}
U_{R1} \longrightarrow -U_{R2},~~~~~~~~
U_{R2} \longrightarrow -U_{R1},~~~~~~~~
u_{R} \longrightarrow u_{R}.
\end{eqnarray}
Then we obtain the following $SU(2)_L$-singlet $T$ parity eigenstates:
\begin{eqnarray}
 U_{L \pm} = \frac{U_{L1} \mp U_{L2}}{\sqrt{2}},~~~~~~
 U_{R \pm} = \frac{U_{R1} \mp U_{R2}}{\sqrt{2}},~~~~~~
 u_R.
\end{eqnarray}
The Yukawa interaction for the top sector is given by
\begin{eqnarray}
&&{\cal L}_{\rm top}
=
-\frac{1}{2 \sqrt{2}} \lambda_1 f \epsilon_{ijk} \epsilon_{xy}
\left[
(\bar{Q}_1)_i (\Sigma)_{jx} (\Sigma)_{ky}
-
(\bar{Q}_2 \Sigma_0)_i (\tilde{\Sigma})_{jx} (\tilde{\Sigma})_{ky}
\right]
u_R \nonumber \\
&&~~~~~~~~~
-\lambda_2 f
\left(
\bar{U}_{L1} U_{R1} + \bar{U}_{L2} U_{R2}
\right)
~+ ~{\rm h.c.},
\end{eqnarray}
where $\tilde{\Sigma} = \Sigma_0 \Omega \Sigma^{\dag} \Omega \Sigma_0$ is the image of $\Sigma$ under $T$ parity, $\lambda_{1,2}$ are the Yukawa coupling constants of the top sector, and $i, j, k$ and $x, y$ run over $1-3$ and $4-5$, respectively. From this Lagrangian we obtain the mass eigenstates of the top quark $t$ and its heavy partners $T_{\pm}$ as
\begin{eqnarray}
&&
\left(
\begin{array}{c}
t_L \\
\left(T_+\right)_L
\end{array}
\right)
=
\left(
\begin{array}{cr}
\cos\theta_L & -\sin\theta_L \\
\sin\theta_L & \cos\theta_L
\end{array}
\right)
\left(
\begin{array}{c}
u_{SM} \\
U_{L+}
\end{array}
\right),~~~~~~~
\left(T_-\right)_L = U_{L-},  \nonumber \\
&&
\left(
\begin{array}{c}
t_R \\
\left(T_+\right)_R
\end{array}
\right)
=
\left(
\begin{array}{cr}
\cos\theta_R & -\sin\theta_R \\
\sin\theta_R & \cos\theta_R
\end{array}
\right)
\left(
\begin{array}{c}
u_{R} \\
U_{R+}
\end{array}
\right),~~~~~~~
\left(T_-\right)_R = U_{R-},
\end{eqnarray}
where $u_{SM}$ is the upper component of the left-hand SM $SU(2)_L$ quark doublet $q_{SM}$. At the ${\cal O}(v^2/f^2)$,
\begin{eqnarray}
\sin\theta_L = x_L \frac{v}{f},~~~~~~
\sin\theta_R
=
\sqrt{x_L}
\left[
1 - \frac{v^2}{f^2} (1 - x_L)
\left(
\frac{1}{2} - x_L
\right)
\right],
\end{eqnarray}
and the masses of $T_+$ and $T_-$ are give by
\begin{eqnarray}
&& m_{T_+}
=
\frac{f}{v} \frac{m_t}{\sqrt{x_L (1 - x_L)}}
\left[
1 + \frac{v^2}{f^2} \left( \frac{1}{3} - x_L (1 - x_L) \right)
\right], \nonumber \\
&& m_{T_-}
=
\frac{f}{v} \frac{m_t}{\sqrt{x_L}}
\left[
1 + \frac{v^2}{f^2} \left( \frac{1}{3} - \frac{1}{2} x_L (1 - x_L) \right)
\right],
\end{eqnarray}
where
\begin{eqnarray}
m_{t}
=
v \sqrt{x_L (1 - x_L) (\lambda_1^2 + \lambda_2^2)}
\left[
1 + \frac{v^2}{f^2} \left( -\frac{1}{3} + \frac{1}{2} x_L (1 - x_L) \right)
\right],~~~~~~
x_L = \lambda_1^2/(\lambda_1^2 + \lambda_2^2).
\end{eqnarray}

\subsection{Related Feynman rules}
\par
The Feynman rules for the vertices in the LHT used in this work are listed in Table \ref{feynrules} \cite{Belyaev:2006jh,Blanke:2006eb,Pukhov:2004ca,THan}, where $V_{Hu}$ and $V_{Hd}$ are two Cabibbo-Kobayashi-Maskawa (CKM)-like unitary mixing matrices for mirror quarks. \footnote {The Feynman rules for the SM Higgs gauge and Yukawa interactions are only valid up to the ${\cal O}(v^2/f^2)$.} These two mirror mixing matrices satisfy $V_{Hu}^{\dag} V_{Hd} = V_{CKM}$, with $V_{CKM}$ being the SM CKM matrix. In the following calculations we take $V_{Hu}$ to be a unit matrix, and therefore we have $V_{Hd} = V_{CKM}$.
\begin{table}[h]
\small
\begin{center}
\begin{tabular}{|c|l||c|l|}
\hline
Vertex & ~~~~~~~~~~~~~~~~~~~~~~Feynman rule & Vertex & ~~~~~~~~~~~~Feynman rule \\
\hline
&&& \\
$\bar{u}_{i-} Z_{H}^{\mu} u_j$
&
$i \left( \frac{g}{2} \cos\theta_H - \frac{g^{\prime}}{10} \sin\theta_H \right) \left( \cos\theta_L \right)^{\delta_{j3}} \gamma^{\mu} P_L \left( V_{Hu} \right)_{ij}$
&
$h \bar{t} t$
&
$-i \frac{g}{2} \frac{m_t}{m_{W}} \left[ 1 - \left( \frac{3}{4} - x_L + x_L^2 \right) \frac{v^2}{f^2} \right]$ \\
&&& \\
$\bar{u}_{i-} Z_{H}^{\mu} T_+$
&
$i \left( \frac{g}{2} \cos\theta_H - \frac{g^{\prime}}{10} \sin\theta_H \right) \left( \sin\theta_L \right) \gamma^{\mu} P_L \left( V_{Hu} \right)_{i3}$
&
$h \bar{T}_+ T_+$
&
$i \frac{g}{2} \frac{m_{T_+}}{m_{W}} \left( x_L - x_L^2 \right) \frac{v^2}{f^2}$ \\
&&& \\
$\bar{d}_{i-} Z_{H}^{\mu} d_j$
&
$i \left( -\frac{g}{2} \cos\theta_H - \frac{g^{\prime}}{10} \sin\theta_H \right) \gamma^{\mu} P_L \left( V_{Hd} \right)_{ij}$
&
$h \bar{u}_{i-} u_{i-}$
&
$i \frac{g}{8} \frac{m_{u_{i-}}}{m_{W}} \frac{v^2}{f^2}$ \\
&&& \\
$\bar{T}_- Z_{H}^{\mu} t$
&
$-i \frac{2}{5} g^{\prime} \sin\theta_H \gamma^{\mu} \Big( \sin\theta_L P_L + \sin\theta_R P_R \Big)$
&
$G^{a \mu} \bar{T}^{\alpha}_{\pm} T^{\beta}_{\pm}$
&
$i g_s \left( T^a \right)_{\alpha \beta} \gamma^{\mu}$ \\
&&& \\
$\bar{T}_- Z_{H}^{\mu} T_+$
&
$i \frac{2}{5} g^{\prime} \sin\theta_H \gamma^{\mu} \Big( \cos\theta_L P_L + \cos\theta_R P_R \Big)$
&
$G^{a \mu} \bar{q}^{\alpha}_{-} q^{\beta}_{-}$
&
$i g_s \left( T^a \right)_{\alpha \beta} \gamma^{\mu}$ \\
&&& \\
$h Z_{H}^{\mu} Z_{H}^{\nu}$
&
$-i g m_W \left( 1 - 2 \tan\theta_W \sin\theta_H - \frac{5}{4} \frac{v^2}{f^2} \right) g^{\mu \nu}$
&& \\
&&& \\
$h Z_H^{\mu} A_H^{\nu}$
&
$-i g^{\prime} m_W \left( 1 + 2 \cot2\theta_W \sin\theta_H - \frac{5}{4} \frac{v^2}{f^2} \right) g^{\mu \nu}$
&& \\
&&& \\
\hline
\end{tabular}
\caption{\label{feynrules}
The related LHT Feynman rules used in this paper, where $P_{L, R} = \frac{1}{2} ( 1 \mp \gamma^5 )$, $q_- = u_-, d_-, c_-, s_-, t_-, b_-$,
$m_W ~(= m_{W_L})$ is the SM $W$-boson mass, $\theta_W$ is Weinberg weak mixing angle, and $i, j = 1, ..., 3$ are flavor indices. }
\end{center}
\end{table}

\vskip 5mm
\section{CALCULATION CONFIGURATION  }
\par
In this work we adopt the 5-flavor scheme and treat the $u$-, $d$-, $c$-, $s$- and $b$-quarks as massless particles.

\par
{\bf A. LO calculation }
\par
We find that the LO cross section for $pp \to Z_HZ_H+X$ process by taking the diagonal CKM matirx, is equal to that by taking a nondiagonal CKM matrix due to the unitary feature of the CKM matrix. Therefore, we can set $V_{CKM}$ to be the unit matrix throughout our calculation without loss of generality. In this case, the $Z_H$-pair production at the LHC is contributed to by the following partonic processes at the lowest order:
\begin{eqnarray}
q(p_1)+\bar{q}(p_2)\to Z_H(p_3)+Z_H(p_4),~~~(q=u,d,c,s,b).
\end{eqnarray}
We present the LO Feynman diagrams for the partonic process $q\bar{q} \to Z_H Z_H$ in Fig.\ref{born.Fig}. The LO cross section for this partonic process is expressed as
\begin{eqnarray}
\label{losigma}
\hat{\sigma}^{(0)}_{q\bar q} \,=\, \frac{1}{2} \frac{(2 \pi
)^4}{2\hat{s}}\int \overline{ \sum} |{\mathcal
M}^{LO}_{q\bar q}|^2 d\Omega_2,&&(q\bar{q}=u\bar{u},d\bar{d},c\bar{c},s\bar{s},b\bar{b})£¬
\end{eqnarray}
where the factor $\frac{1}{2}$ arises from the two identical particles in the final state, $\sqrt{\hat{s}}$ is the colliding energy in the partonic center-of-mass system (c.m.s), and ${\cal M}^{LO}_{q\bar q}$ is the LO amplitude for $q\overline q \to Z_H Z_H$. The summation is taken over the spins of the final state, and the bar over the summation represents averaging over the spins and colors of the initial state. The phase space element of the two-body final states is expressed as
\begin{eqnarray}
d\Omega_2 \,=\, \delta^{(4)}(p_1+p_2-p_3-p_4) \, \frac{d^3
\vec{p}_3}{2E_3 (2\pi)^3}\, \frac{d^3 \vec{p}_4}{2E_4(2\pi)^3}.
\end{eqnarray}
\begin{figure}[tb]
\centering
\includegraphics[scale=1.0]{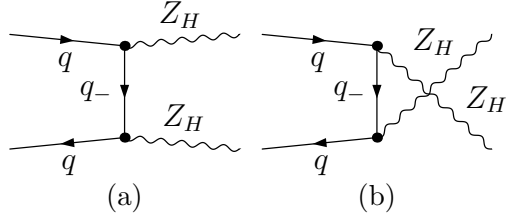}
\caption{The LO Feynman diagrams for the partonic process $q\bar{q}\to Z_HZ_H$.}
\label{born.Fig}
\end{figure}

\par
The cross section for the parent process $pp  \to q \bar{q} \to Z_H Z_H+X$ at the LO can be obtained by the following convolution:
\begin{align}
\label{diffhadr}
\sigma^{(0)}(pp \to q\bar{q} \to Z_H Z_H+X) \, = \,
\sum_{q\bar q} \int_{\tau_{0}}^{1}d\tau \, \int_{\tau}^{1} \frac{dx}{x}\,
\left[ f_{q/P_1}(x,\mu_{f})\, f_{\bar q/P_2}\!\left(\frac{\tau}{x},\mu_{f}\right) \,
 +\, (P_1 \leftrightarrow P_2) \right] \,
\hat{\sigma}^{(0)}_{q \bar q}(\tau) \, ,
\end{align}
where the summation runs over all possible initial $q\bar q$ states (i.e., $q\bar q=u\bar u$, $d\bar d$, $c\bar c$, $s\bar s$, $b\bar b$), $f_{q/P_i}(x,\mu_{f})$ denotes the parton distribution function (PDF) at the scale $\mu_f$ of  parton $q$ with momentum fraction $x$ in the proton, $\tau$ is the ratio between the squared c.m.s energies of the partonic and hadronic processes, i.e., $\tau=\hat{s}/s$, and the kinematical production threshold $\tau_{0}=4m_{Z_H}^2/s$.

\par
{\bf B. Virtual and real corrections }
\par
The next-to-leading order (NLO) QCD corrections to the parent process $pp  \to q \bar{q} \to Z_H Z_H+X$ involve ultraviolet (UV), soft and collinear infrared (IR) singularities. We employ the dimensional regularization (DR) method in $D\,=\, 4-2\epsilon$ dimensions to isolate the relevant singularities. The UV divergences are removed by the renormalization procedure. The soft singularities vanish after summing up the virtual correction and the real gluon bremsstrahlung contribution. The collinear singularities are partially canceled by the real light-quark/gluon emission contributions, and the remaining collinear singularities are absorbed by the PDF counterterms.

\par
We define the relevant renormalization constants as
\begin{equation}
\psi_{q,L,R}^0 \, =\, \left( 1 + \frac{1}{2}\delta Z_{q,L,R} \right) \psi_{q,L,R} \, ,
~~~~\psi_{q_-,L,R}^0 \, =\, \left( 1 + \frac{1}{2}\delta Z_{q_-,L,R} \right) \psi_{q_-,L,R} \, ,~~~~
m_{q_-}^0 \,=\, m_{q_-} + \delta m_{q_-} \; ,
\end{equation}
where $\psi_q$, $\psi_{q_-}$, and $m_{q_-}$ are the light-quark field, $T$-odd heavy quark field, and $q_-$ mass, respectively, and the  superscript 0 indicates the corresponding bare quantity. In the on shell renormalization scheme, these renormalization constants are expressed as
\begin{align}
\delta Z_{q,L}\, &= \delta Z_{q,R}= \, -\frac{\alpha_s(\mu_r)}{3\pi} \left[\Delta_{UV}-\Delta_{IR}\right] \, ,
  \nonumber \\
\delta Z_{q_-,L}\, &= \delta Z_{q_-,R}= \, -\frac{\alpha_s(\mu_r)}{3\pi} \left[\Delta_{UV}+2\Delta_{IR} +4 + 3 \ln{\frac{\mu^2_r}{m^2_{q_-}}} \right] \, ,
  \nonumber \\
\frac{\delta m_{q_-}}{m_{q_-}} \, &= \, -\frac{\alpha_s(\mu_r)}{3\pi}\left[ 3\left(\Delta_{UV}+
     \ln{\frac{\mu^2_r}{m^2_{q_-}}} \right)+4 \right] \; ,
\end{align}
with the definitions of $\Delta_{UV}=\frac{1}{\epsilon_{UV}}-\gamma_E + \ln (4\pi)$
and $\Delta_{IR}=\frac{1}{\epsilon_{IR}}-\gamma_E + \ln (4\pi)$.

\par
The one-loop Feynman diagrams for the partonic process $q\bar{q}\to Z_HZ_H$ are shown in Fig.\ref{vir.Fig}. The UV singularities in the one-loop amplitudes are canceled by the counterterm of the total amplitude, which is the sum of all counterterms for the related light-quark fields, propagators, and vertices, expressed as
\begin{align}
{\cal M}^{CT}_{q\bar{q}}\, = \, &
\delta m_{q_-}\,\left(\,
{\cal M}^{(t)}_{q\bar{q}}\mid_{\frac{i}{\not{p}_1-\not{p}_3 -m_{q_-}}
\rightarrow \frac{i}{(\not{p}_1-\not{p}_3-m_{q_-})^2}}
\,+\,
{\cal M}^{(u)}_{q\bar{q}}\mid_{\frac{i}{\not{p}_1-\not{p}_4-m_{q_-}}
\rightarrow \frac{i}{(\not{p}_1-\not{p}_4-m_{q_-})^2}}
\;\right)
 \nonumber  \\
 & +\,
 \delta Z_{q,L} \,{\cal M}^{LO}_{q\bar{q}}\, ,
\end{align}
where ${\cal M}^{LO}_{q\bar{q}}={\cal M}^{(t)}_{q\bar{q}}+{\cal M}^{(u)}_{q\bar{q}}$; the two amplitudes in parentheses represent those obtained by applying the replacement of ${\frac{i}{\not{p}_1-\not{p}_3-m_{q_-}}\rightarrow \frac{i}{(\not{p}_1-\not{p}_3-m_{q_-})^2}}$ in the $t$-channel LO amplitude and the replacement of ${\frac{i}{\not{p}_1-\not{p}_4-m_{q_-}}\rightarrow \frac{i}{(\not{p}_1-\not{p}_4-m_{q_-})^2}}$ in the $u$-channel LO amplitude, separately.
\begin{figure}[tb]
\centering
\includegraphics[scale=0.7]{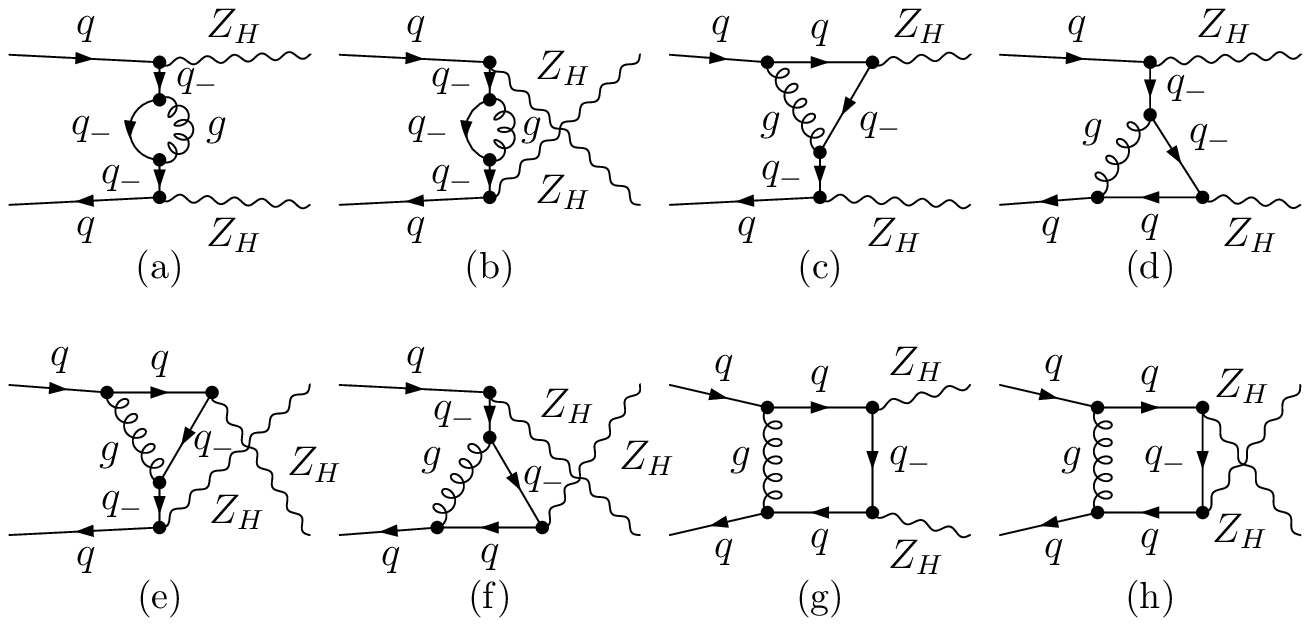}
\caption{ The one-loop Feynman diagrams for the partonic process $q\bar{q}\to Z_HZ_H$.}
\label{vir.Fig}
\end{figure}
\begin{figure}[tb]
\centering
\includegraphics[scale=0.7]{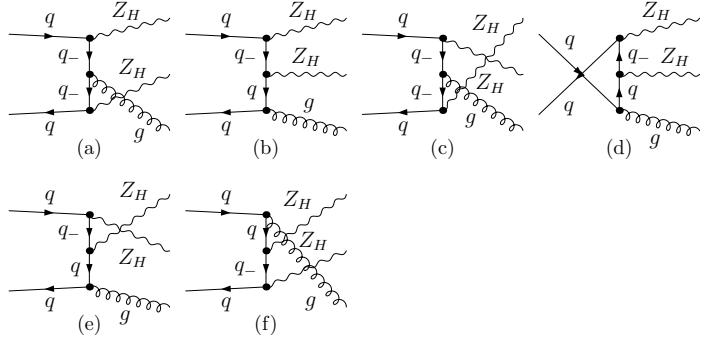}
\caption{The Feynman diagrams for the real gluon emission process $q\bar{q}\to Z_HZ_H +g$. }
\label{rg.Fig}
\end{figure}

\par
The diagrams for the real gluon emission contribution are shown in Fig.\ref{rg.Fig}. To isolate the soft and collinear singularities in the real gluon emission $q(p_1)\bar{q}(p_2) \to Z_H(p_3) Z_H(p_4)+g(p_5)$, we employ the two-cutoff phase space slicing (TCPSS) method \cite{Harris:2001sx}. An arbitrary soft cutoff $\delta_{s}$ is introduced to separate the real gluon emission subprocess phase space into two regions, soft gluon ($E_{5} \leq \delta_{s} \sqrt{\hat{s}}/2$) and hard gluon regions ($E_{5} > \delta_{s} \sqrt{\hat{s}}/2$). Furthermore, another cutoff $\delta_{c}$ is introduced to decompose the hard gluon region into a hard collinear ($HC$) region ($\hat{s}_{15}$ or $\hat{s}_{25} \leq \delta_{c}\hat{s}$)) and a hard noncollinear ($\overline{HC}$) region ($\hat{s}_{15}$ and $\hat{s}_{25} > \delta_{c}\hat{s}$) with $\hat{s}_{ij} = (p_i+p_j)^2$.
The soft gluon emission plus the virtual contribution to the cross section for the parent process $pp\to q\bar {q} \to Z_HZ_H+X$ has the form 
\begin{align}
\sigma^{V}+\sigma_{(g)}^{S} \,&= \,
\sum_{q\bar q} \int_{\tau_{0}}^{1}d\tau \, \mathcal{L}_{q\bar q}(\tau)\,
\left[\hat{\sigma}^{V}(\tau)+\hat{\sigma}^{S}_{(g)}(\tau)\right] \, .
\end{align}
The soft contribution is written as
\begin{eqnarray}
\label{soft}
\hat{\sigma}^{S}_{(g)}(\tau)\,=\,
\hat{\sigma}^0_{q \bar q}(\tau) \left[\frac{\alpha_s}{2 \pi}
\frac{\Gamma(1-\epsilon)}{\Gamma(1-2 \epsilon)}\left(\frac{4 \pi
\mu_r^2}{\hat{s}}\right)^{\epsilon}\right]
\left(\frac{A^S_2}{\epsilon^2}+\frac{A^S_1}{\epsilon}+A^S_0 \right),
\end{eqnarray}
with
\begin{eqnarray}
A^S_2&=& 2 C_F,~~~A^S_1 = - 4 C_F \ln \delta_s, ~~~A^S_0 = 4 C_F
\ln^2 \delta_s,
\end{eqnarray}
where $C_F=4/3$ and the parton luminosity is expressed as
\begin{align}
\label{luminosity}
\mathcal{L}_{q\bar q}(\tau) \, = \, \int_{\tau}^{1} \frac{dx}{x}\,
\left[ f_{q/P_1}(x,\mu_{f})\, f_{\bar q/P_2}\!\left(\frac{\tau}{x},\mu_{f}\right) \,
 +\, (P_1 \leftrightarrow P_2) \right] \, .
\end{align}
Factorizing the relevant collinear singularities into the PDFs, the real gluon emission correction over the hard collinear region is expressed as
\begin{align}
\sigma_{(g)}^{C}\,  = & \,
\sum_{q\bar q}
\Biggl\{
\int_{\tau_{0}}^{1}d\tau\,\int_{\tau}^{1}\frac{dx}{x}\,\int_{x}^{1-\delta_s}\frac{dz}{z}
\left[
P_{qq}(z)\ln\left(\delta_c\frac{1-z}{z}\frac{\hat{s}}{\mu_f^2}\right)-P_{qq}^{'}(z)
\right]
{\mathcal L}^{coll}_{q\bar{q}}(\tau,x,z)
\nonumber \\  &
\,+\,
\int_{\tau_{0}}^{1}d\tau \,
\left[ 2\left( \frac{A_1^{sc}}{\epsilon}+A_0^{sc}\right)
{\mathcal L}_{q\bar{q}}(\tau)
\right]
\Biggr\}
\hat{\sigma}^{(0)}_{q\bar{q}}(\tau)
\left[
\frac{\alpha_s}{2\pi}\frac{\Gamma\left(1-\epsilon\right)}{\Gamma\left(1-2\epsilon\right)}
\left(\frac{4\pi\mu_r^2}{\hat{s}}\right)^{\epsilon}
\right],
\end{align}
with
\begin{align}
{\mathcal L}_{q\bar{q}}^{coll}(\tau,x,z)\,=&\,
f_{\bar{q}/P_1}\left(\frac{\tau}{x},\mu_f \right)f_{q/P_2}\left(\frac{x}{z},\mu_f\right)\,+\,
f_{\bar{q}/P_1}\left(\frac{x}{z},\mu_f \right)f_{q/P_2}\left(\frac{\tau}{x},\mu_f \right)
+(P_1 \leftrightarrow P_2)\, , \nonumber \\
P_{qq}(z)\,=&\,C_F\frac{1+z^2}{1-z}\, ,~~~~~~~~
P_{qq}^{'}(z)\,=\, - C_F \left( 1-z \right) \, , \nonumber \\
A_1^{sc}\,=&\,C_F\left( 2\ln \delta_s + 3/2 \right) \, , ~~~~~~~~
A_0^{sc}\,=\, A_1^{sc}\ln\left(\frac{\hat{s}}{\mu_f^2} \right) \, .
\end{align}
Finally, the residual hard noncollinear cross section $\sigma^{NC}_{(g)} $ integrated over the phase space outside the soft and hard collinear region is IR finite and can be evaluated in four dimensions by using Monte Carlo technique. The summation of these three parts of real gluon bremsstrahlung contributions and the virtual correction is independent of the TCPSS cutoffs in the range of $\delta_s\in [1\times 10^{-4}, ~1\times 10^{-6}]$ and $\delta_c=\delta_s/100$, which is verified with high precision in our numerical calculation. Therefore, we did not use the subtraction method (like the Catani-Seymour or Frixione-Kunszt-Signer scheme) \cite{CS,FKS}.

\par
The light-quark-gluon scattering subprocesses $q[\bar{q}]g \to Z_H Z_H+q[\bar{q}]$ $(q=u,d,c,s,b)$ also contribute to the $\mathcal O(\alpha_s)$ corrections to the $pp \to q\bar{q} \to Z_H Z_H +X$ process. The real antiquark emission subprocess is similar to the real quark emission subprocess, so we only show the calculation of real quark emission subprocesses. Their tree-level Feynman diagrams are shown in Figs.\ref{fig5a}(a)-\ref{fig5a}(f). In our chosen parameter space in this work, each of the diagrams in Figs.\ref{fig5a}(c)-\ref{fig5a}(f) contains a possible $q_-$-resonance propagator due to the mass of the $T$-odd quark being larger than that of $Z_H$, while the diagrams in Figs.\ref{fig5a}(a)-\ref{fig5a}(b) do not include a $q_-$-resonance propagator. As a matter of bookkeeping, the $q_-$-resonance production mechanism is more intuitively interpreted as the on shell $Z_H q_-$ production with subsequent decay of $q_- \to Z_H q$. In order to avoid double counting and not to artificially spoil the convergence of the perturbation, we subtract the contributions of the $q[\bar{q}] g \to Z_H q_- \to Z_HZ_H + q[\bar{q}]$ subprocesses that are mediated by on shell $Z_H$ and $q_-$ from the associated NLO QCD corrections. There are several approaches that can be used to carry out this subtraction. In this work, we employ the diagram subtraction scheme \cite{Frixione:2008yi} that implements a local and gauge-invariant subtraction term. In applying this scheme, we have to use the total decay width of $q_-$-quark $\Gamma_{q_-}$ in the quantitative computation of the real light-quark bremsstrahlung contribution, which will modify their collinear limits, and therefore would spoil the local cancellation of collinear singularities. In our work, we take the pragmatic prescription that has been extensively used at the large electron positron collider in handling the resonance decays \cite{Frixione:2008yi}. Namely, only in the resonant propagator the $\Gamma_{q_-}\neq 0$ is applied and the rest nonresonant propagators $\Gamma_{q_-}=0$. The subtracted real light-quark emission contribution $\hat{\sigma}_{(q)}^{R}$ is given by
\begin{align}
\hat{\sigma}_{(q)}^{R}  \, \sim \,&  \int \,
\overline{\sum} \Biggl[
\left|{\mathcal M}\right|^2
-\frac{m_{q_-}^2\Gamma_{q_-}^2}{\left(s_{45}-m_{q_-}^2\right)^2+m_{q_-}^2\Gamma_{q_-}^2}\,\left|{\mathcal M}_{R}^{(c+d)}(s_{45}=m_{q_-}^2)
\right|^2
\nonumber \\ &
-\frac{m_{q_-}^2\Gamma_{q_-}^2}{\left(s_{35}-m_{q_-}^2\right)^2+m_{q_-}^2\Gamma_{q_-}^2}\,\left|{\mathcal M}_{R}^{(e+f)}(s_{35}=m_{q_-}^2)
\right|^2 \Biggl] d\Omega_3,
\end{align}
where the summation is taken over the spins and colors of final state, and the bar over the summation means taking averages over initial spin and color states; and $d\Omega_3$ is the three-body phase space element; ${\mathcal M}={\mathcal M}_{NR}+{\mathcal M}_{R}$ corresponds to the total amplitude, and ${\mathcal M}_{NR}={\mathcal M}^{(a)}+{\mathcal M}^{(b)}$ is the amplitude for the nonresonance diagrams of Figs.\ref{fig5a}(a)-\ref{fig5a}(b), ${\mathcal M}_{R}={\mathcal M}^{(c)}+{\mathcal M}^{(d)}+{\mathcal M}^{(e)}+{\mathcal M}^{(f)}$ is for all the diagrams in Figs.\ref{fig5a}(c)-\ref{fig5a}(f) containing a $q_-$-resonance propagator. There, ${\mathcal M}_{R}^{(c+d)}={\mathcal M}^{(c)}+{\mathcal M}^{(d)}$, ${\mathcal M}_{R}^{(e+f)}={\mathcal M}^{(e)}+{\mathcal M}^{(f)}$, $s_{45}=(p_4+p_5)^2$, $s_{35}=(p_3+p_5)^2$, and in the amplitudes ${\mathcal M}_{R}^{(c)}$, ${\mathcal M}_{R}^{(d)}$, ${\mathcal M}_{R}^{(e)}$, and ${\mathcal M}_{R}^{(f)}$ the resonant $q_-$-propagators are all in the form of $\frac{i}{\slashed{p}-m_{q_-}+i\Gamma_{q_-}}$ with $\Gamma_{q_-}\neq 0$.
\begin{figure}
\centering
\includegraphics[scale=0.7]{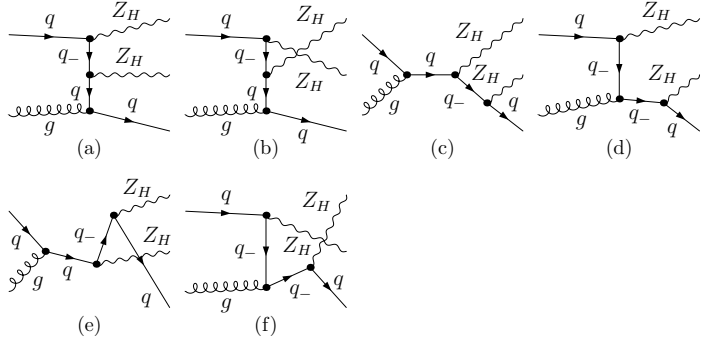}
\caption{The tree-level Feyman diagrams for the $q g \to Z_HZ_H+q$ partonic process. The diagrams of Figs.\ref{fig5a}(a)-\ref{fig5a}(b) do not include a $q_-$-resonance propagator, while each of the diagrams in Figs.\ref{fig5a}(c)-\ref{fig5a}(f) contains a possible $q_-$-resonance propagator.}
\label{fig5a}
\end{figure}

\par
We adopt again the TCPSS method to isolate the collinear singularities in the calculation of the real light-quark emission correction. In adopting this method, the real light-quark emission contribution can be separated into collinear and noncollinear parts, i.e., $\sigma_{(q)}^{R} =\sigma_{(q)}^{C}+\sigma_{(q)}^{NC}$. Here, we meet only initial state collinear divergences in calculation. Factorizing and absorbing the relevant collinear singularities into the PDFs, the real light-quark emission correction over the collinear region can be written as
\begin{align}
\sigma_{(q)}^{C}  \, =&\,
\sum_{q} \int_{\tau_{0}}^1d\tau \, \int_{\tau}^1\frac{dx}{x} \, \int_{x}^1 \frac{dz}{z} \,
\mathcal{L}_{qg}^{coll}(\tau,x,z) \,
\hat\sigma^{(0)}_{q\bar q}(\tau)  \nonumber \\
& \frac{\alpha_s}{2\pi}\frac{\Gamma\left(1-\epsilon\right)}{\Gamma\left(1-2\epsilon\right)}
\left(\frac{4\pi\mu_r^2}{\hat{s}}\right)^{\epsilon} \left[
       P_{qg}(z)\ln \left(\delta_c\frac{1-z}{z}\frac{\hat{s}}{\mu_f^2} \right)
       - P_{qg}^{'}(z)  \right] \, ,
\end{align}
with
\begin{align}
{\mathcal L}_{qg}^{coll}(\tau,x,z) \,=& \,
f_{q/P_1}\left(\frac{\tau}{x},\mu_f\right) \,f_{g/P_2}\left(\frac{x}{z},\mu_f\right)
+\left( P_1 \leftrightarrow P_2 \right)\, , \nonumber \\
P_{qg}(z)\,=&\, \frac{1}{2}\left[ z^2+(1-z)^2 \right] \, , ~~~~~~~~~~~~
P_{qg}^{'}(z)\,=\,-z \left( 1-z \right) \, .
\end{align}
The independence on the TCPSS cutoff parameter $\delta_c$ of the real light-quark emission contribution is verified too.

\par
{\bf C. $gg$-fusion correction }
\par
The NLO QCD correction to the partonic process $q\bar q \to Z_HZ_H$ is of the order of $\alpha_{ew}^2\alpha_{s}$, while the \ggZHZH subprocess is induced via one-loop diagrams at the lowest order of $\alpha_{ew}^2\alpha_{s}^2$. Although the lowest order contribution of the gluon-gluon fusion subprocess is $\alpha_{s}$ order higher than the previous one, the contribution from the \ggZHZH subprocess at the $\sqrt{s}=14~{\rm TeV}$ LHC might be non-negligible due to the high gluon luminosity in proton. The representative Feynman diagrams for the \ggZHZH partonic process are depicted in Fig.\ref{fig5b}. The total one-loop amplitude ${\cal M}^{1-loop}_{gg}$ for this partonic process is UV- and IR finite, the cross section at the lowest order, $\hat{\sigma}_{gg}$, can be expressed as
\begin{eqnarray}
\hat{\sigma}_{gg} \,=\, \frac{1}{2} \frac{(2 \pi
)^4}{2\hat{s}}\int \overline{ \sum} |{\mathcal
M}^{1-loop}_{gg}|^2 d\Omega_2 \, ,
\end{eqnarray}
and the hadronic cross section for the parent process $pp  \to gg \to Z_H Z_H+X$ at the lowest order can be obtained by the convolution of
\begin{align}
\label{diffhadrgg}
\sigma\left(pp \to gg \to Z_H Z_H +X\right) \, = \,
 \int_{\tau_{0}}^{1}d\tau \, \int_{\tau}^{1} \frac{dx}{x}\,
 f_{g/P_1}(x,\mu_{f})\, f_{ g/P_2}\!\left(\frac{\tau}{x},\mu_{f}\right) \,
\hat{\sigma}_{gg}(\tau) \, ,
\end{align}
where we adopt the same notations as in Eqs.(\ref{losigma}) and (\ref{diffhadr}), and $f_{g/P_i}(x,\mu)$ denotes the gluon PDF in proton.
\begin{figure}
\begin{center}
\includegraphics[scale=0.7]{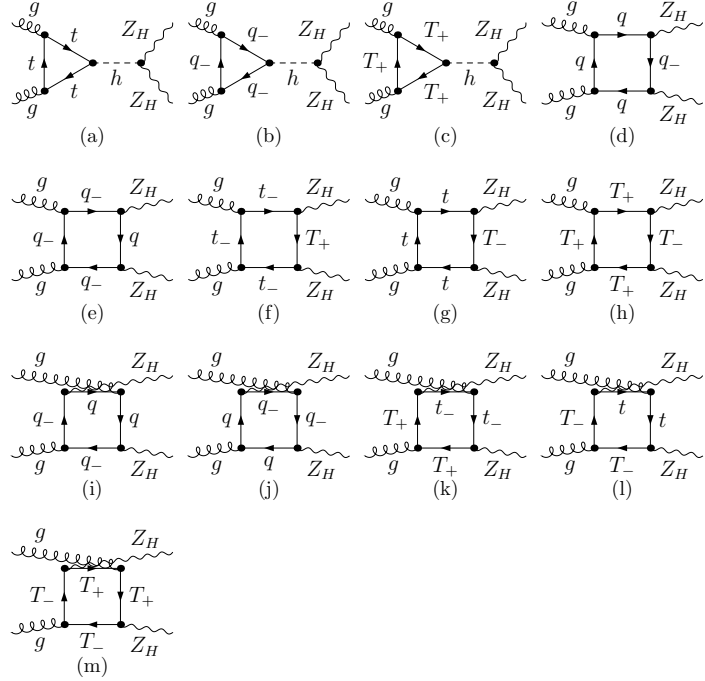}
\caption{ \label{fig5b} The representative QCD one-loop Feynman diagrams for the partonic process \ggZHZH, where $q=u,d,c,s,b,t$ and $q_-=u_-,d_-,c_-,s_-,b_-,t_-$. }
\end{center}
\end{figure}

\par
From the above calculation strategy, we can obtain the total QCD correction to the integrated cross section for the $pp \to q\bar q \to Z_HZ_H+X$ process as
\begin{eqnarray}\label{TotalSigmaCorr}
&&\Delta \sigma_{QCD}(pp \to Z_H Z_H+X)=\Delta\sigma^{(2)}+\Delta\sigma^{(3)}  \nb \\
&=& \sigma^{V}+\sigma_{(g)}^{S}+\sigma_{(g)}^{C}+\sigma_{(q)}^{C}+\sigma_{(g)}^{NC}+\sigma_{(q)}^{NC}+\sigma(pp \to gg \to Z_H Z_H+X)  \nb \\
&=&\Delta\sigma_{NLO}(pp \to q\bar{q} \to Z_H Z_H+X)
+\sigma(pp \to gg \to Z_H Z_H+X).
\end{eqnarray}
And the total QCD corrected integrated cross section for the parent $pp \to Z_HZ_H+X$ process, defined as involving the pure NLO QCD correction and the gluon-gluon fusion correction, can be expressed by
\begin{eqnarray}\label{QCD-sigma}
&&\sigma_{QCD}(pp \to Z_H Z_H+X)= \sigma^{(0)}(pp \to q\bar{q} \to Z_H Z_H+X)+\Delta \sigma_{QCD}(pp \to Z_H Z_H+X)  \nb \\
&=& \sigma^{(0)}(pp \to q\bar{q} \to Z_H Z_H+X)+\Delta\sigma_{NLO}(pp \to q\bar{q} \to Z_H Z_H+X)+\sigma(pp \to gg \to Z_H Z_H+X). \nb \\
\end{eqnarray}

\vskip 5mm
\section{RESULTS AND DISCUSSIONS }
\par
\subsection{General input parameters}
\par
We take $\alpha_{ew}(m_Z^2)^{-1}=127.944$, $m_W=80.385~{\rm GeV}$, $m_Z=91.1876~{\rm GeV}$, $m_{t}=173.5~{\rm GeV}$, \cite{Beringer:1900zz} and $m_h=125~{\rm GeV}$. For simplicity we set the factorization and renormalization scales as equal ($\mu=\mu_r=\mu_f$) and define the central scale as $\mu_0=m_{Z_H}$. We adopt the CTEQ6L1 and CTEQ6M PDFs \cite{Pumplin:2002vw} for the LO and QCD higher order calculations, separately. The strong coupling constant $\alpha_s(\mu)$ is determined by the QCD parameter $\Lambda_5^{LO} = 165~{\rm MeV}$ for the CTEQ6L1 and $\Lambda_5^{\overline{MS}} = 226~{\rm MeV}$ for the CTEQ6M, respectively. We assume the branch ratio of the LHT Higgs boson decay $h \to b \bar{b}$ to be the same as that in the SM, and obtain $Br(h \to b \bar{b}) = 60.70\%$ by adopting the program HDECAY for SM Higgs boson decays \cite{Djouadi:1997yw} with the input SM parameters from Ref.\cite{Beringer:1900zz}. An up-to-date overview of the LHT constraints using the latest results from the $8~{\rm TeV}$ run at the LHC has been presented in Ref.\cite{Reuter:2013iya}. In our following calculations we take the related LHT parameters as $\kappa=1$, $x_L = 1/2$, $f=800~{\rm GeV}$, which are within the surviving LHT parameter space, if there is no other statement. By adopting the above input parameters, we get the branch ratio $Br(Z_H \to A_H h)=100\%$ \cite{Reuter:2013iya}, and all the partial decay widths of the $T$-odd quark are numerically obtained by applying the expressions presented in Ref.\cite{SongMing:2012gb}.

\par
\subsection{Integrated cross sections }
\par
{\bf A. Renormalization/factorization scale dependence  }
\par
The stabilization of the renormalization/factorization scale dependence is one of the main reasons for the requirement of higher order prediction in hadron collider physics. In Fig.\ref{mu.Fig} we depict the LO, pure NLO QCD, total QCD corrected integrated cross sections, and the corresponding $K$-factors, defined as $K\equiv \sigma_{NLO,QCD}/\sigma_{\text{LO}}$, as functions of the renormalization/factorization scale $\mu$ for the process $pp \to Z_H Z_H + X$ at the $\sqrt{s}=14~{\rm TeV}$ LHC. The dashed, solid and long-dashed curves are for the LO, pure NLO QCD, and total QCD corrected integrated cross sections  ($\sigma_{LO}$, $\sigma_{NLO}$, $\sigma_{QCD}$), respectively, where the total QCD correction includes both the pure NLO QCD and the $gg$-fusion contributions. From Fig.\ref{mu.Fig} we can read out that at the central scale $\mu=\mu_0=m_{Z_H}$ the LO and pure NLO QCD corrected integrated cross sections are $\sigma_{LO}=4.1285_{-0.6012}^{+0.8022}~[fb]$ and $\sigma_{NLO}=5.408_{-0.457}^{+0.562}~[fb]$ with the scale running in the range of $0.2\mu_0 \le \mu \le 5\mu_0$. The scale uncertainty describes the missing higher order corrections estimated via scale variations, and here we define the relative scale uncertainty as $\eta= \left[max(\sigma(\mu))-min(\sigma(\mu))\right]/\sigma(\mu_0)$ with $\mu \in [0.2\mu_0,~5\mu_0]$. We obtain that the pure NLO QCD correction reduces the scale uncertainty from $33.99\%$ (LO) to $18.84\%$ (NLO). The total QCD corrected integrated cross section is $\sigma_{QCD}=5.549_{-0.527}^{+0.748}~[fb]$ and the corresponding relative scale uncertainty is $22.98\%$. Therefore, we can conclude that the relative scale uncertainty within the scale variation of $0.2 \mu_0 < \mu < 5 \mu_0$ is slightly improved by the pure NLO QCD correction, while the scale stabilization deteriorates a little when the $gg$-fusion contribution is included. It also shows that the $gg$-fusion contribution is significant and should be considered together with the pure NLO QCD correction to the $pp \to q\bar{q} \to Z_H Z_H + X$ process in precision prediction. We can read out that the $K$-factor for including only the pure NLO QCD correction varies from $1.21$ to $1.40$ in the plotted scale range, while if taking account of the $gg$-fusion contribution together with the pure NLO QCD correction the $K$-factor varies from $1.28$ to $1.42$. We find also from Fig.\ref{mu.Fig} that the $\mu$ dependence of the LO curve is much stronger than the corresponding ones shown in Refs.\cite{Lazopoulos} and \cite{Dao} for the $ZZZ$ and $WWZ$ productions at the LHC separately (in Figs.4 in both of the references). That might be owing to the production threshold for $Z_H$-pair production being much larger than those for the $ZZZ$ and $WWZ$ productions, which makes the PDF $f_{i/P}(x,\mu)$ contribution to the $pp \to Z_HZ_H+X$ process become more sensitive to the scale $\mu$.

\par
In Table \ref{tab2}, we list the LO, pure NLO QCD, total QCD corrected integrated cross sections and the corresponding $K$-factors by taking fixed scales and transverse energy dependent scales. The central value of the transverse energy dependent scale is defined as $\mu_1 =E_T/2=\frac{1}{2} \sum\limits_i E_{T,i}$, where $E_{T,i}=\sqrt{p_{T,i}^2+m_i^2}$ and the summation is taken over the transverse energies of all final particles. We see from the table that the relative discrepancy between the pure NLO QCD corrected cross sections obtained by taking $\mu=\mu_0$ and $\mu=\mu_1$ is about $3\%$, which is nearly the same as that between the LO cross sections by taking these two scale choices.
\begin{table}
\begin{center}
\begin{tabular}{ c|c|c|c|c}
\hline \hline
$\mu$   & $\sigma_{LO}[fb]$ & $\sigma_{NLO}[fb]$ & $\sigma_{QCD}[fb]$ & $K$  \\
\hline \hline
  $0.5 \mu_0$  & 4.4446(2) & 5.629(11) & 5.828(11) & 1.31    \\
  $~~~\mu_0$   & 4.1285(2) & 5.408(9)  & 5.549(9) & 1.34    \\
  $~~2 \mu_0$  & 3.8493(2) & 5.200(8)  & 5.303(9) & 1.38    \\
\hline \hline
  $0.5 \mu_1$   & 4.2883(3) & 5.437(13) & 5.594(13)  & 1.30    \\
  $~~~\mu_1$    & 3.9935(3) & 5.244(12) &  5.359(12) & 1.34    \\
  $~~2 \mu_1$   & 3.7317(3) & 5.059(11) &  5.144(11) &  1.38   \\
\hline
\end{tabular}
\caption{The LO, pure NLO QCD, total QCD corrected integrated cross sections ($\sigma_{LO}$, $\sigma_{NLO}$, $\sigma_{QCD}$) at the $14~{\rm TeV}$ LHC and the $K$-factors for including the total QCD correction ($K=\sigma_{QCD}/\sigma_{LO}$) with different scale choices, where $\mu_0=m_{Z_H}$ and $\mu_{1}=E_T/2$.  }
\label{tab2}
\end{center}
\end{table}
\begin{figure}
\begin{center}
\includegraphics[scale=0.6]{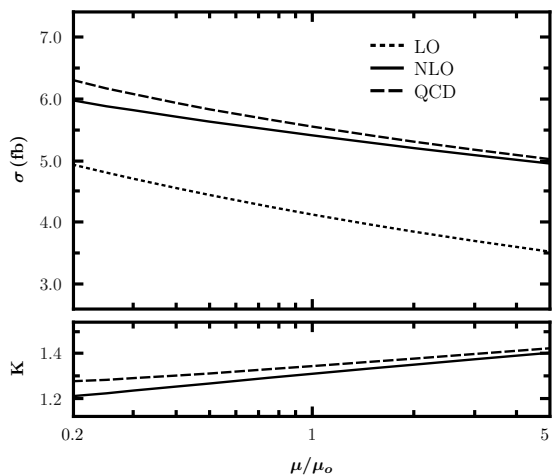}
\caption{The renormalization/factorization scale dependence of the LO, pure NLO QCD, total QCD corrected total cross sections, and the corresponding $K$-factors for the process $pp \to Z_H Z_H+X$ at the $14~{\rm TeV}$ LHC with $\mu_0=m_{Z_H}$.  }
\label{mu.Fig}
\end{center}
\end{figure}

\par
{\bf B. Dependence on symmetry breaking scale $f$ }
\par
The LO, total QCD corrected integrated cross sections and the corresponding $K$-factor for the $pp \to Z_H Z_H + X$ process at the $14~{\rm TeV}$ LHC as functions of the collective symmetry breaking scale $f$ are depicted in Fig.\ref{frun.Fig}. From this figure we can see that the LO and total QCD corrected integrated cross sections diminish dramatically as the increment of $f$. This behavior is due to the $T$-odd $Z_H$ gauge boson becoming heavier as the increment of the symmetry breaking scale $f$, and it makes the phase space of final state smaller. The $K$-factor in the plotted $f$ range varies from $1.24$ to $1.42$. The decrement of the $K$-factor is mainly due to the decrease of the final state phase space, which causes the contribution to $K$-factor from the real gluon/light-quark emission $(2 \to 3)$ processes becoming smaller.
\begin{figure}
\begin{center}
\includegraphics[scale=0.6]{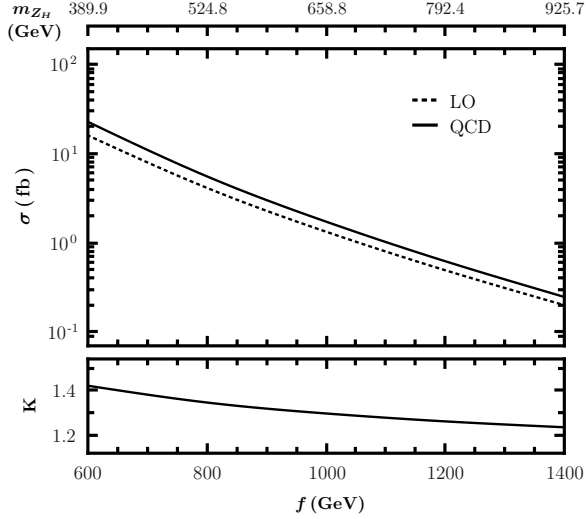}
\caption{The LO, total QCD corrected integrated cross sections, and the corresponding $K$-factor for the process $pp \to Z_H Z_H+X$ at the $14~{\rm TeV}$ LHC as the functions of the symmetry breaking scale $f$.  }
\label{frun.Fig}
\end{center}
\end{figure}

\par
\subsection{Differential cross sections }
\par
{\bf A. Differential cross sections for $pp\to Z_H Z_H \to A_H A_H hh+X$ }
\par
With our choosing typical LHT parameters we have $m_{Z_H}=524.8~{\rm GeV}$, $m_{A_H}=119.7~{\rm GeV}$, and the total decay width $\Gamma_{Z_H}^{total}=5.155\times 10^{-2}~{\rm GeV}$ by using following formula:
\begin{eqnarray}
\Gamma_{Z_H}^{total} ~=~ \Gamma_{Z_H\to A_Hh} = \frac{\left(g^{Z_H}\right)^2}{192\pi} \frac{m_{Z_H}}{m_{A_H}^2}   \sqrt{\lambda }
\left[ \left( 1 -r_h + r_{A_H} \right)^2 + 8 r_{A_H} \right],
\end{eqnarray}
where $g^{Z_H}=g^{\prime} m_W \left( 1 + 2 \cot2\theta_W \sin\theta_H - \frac{5}{4} \frac{v^2}{f^2} \right)$, $r_h=\frac{m_h^2}{m_{Z_H}^2}$, $r_{A_H}=\frac{m_{A_H}^2}{m_{Z_H}^2}$ and $\lambda=1+r_h^2+ r_{A_H}^2-2 r_h r_{A_H}-2r_h-2r_{A_H}$ . Then we obtain $\frac{\Gamma_{Z_H}^{total}}{m_{Z_H}}=9.82\times 10^{-5}$. It shows that it is completely justified to adopt the narrow width approximation (NWA) in choosing our parameter space. In the following we use the Monte Carlo method and the NWA to study the kinematic distributions of the final products for the $pp\to Z_H Z_H \to A_H A_H hh+X$ process with the inclusive event scheme at the $\sqrt{s}=14~{\rm TeV}$ LHC.

\par
The LO, total QCD corrected distributions of the leading Higgs boson transverse momentum $p_T(h_1)$ and the corresponding $K$-factor for the $pp\to Z_H Z_H \to A_H A_H hh+X$ process are depicted in Fig.\ref{pth.Fig}. From Fig.\ref{pth.Fig} we find that the total QCD correction obviously enhances the LO differential cross section in the whole plotted $p_T(h_1)$ range, and the distribution peaks are located in the vicinity of $p_T(h_1) \sim 270~{\rm GeV}$. The corresponding $K$-factor varies from $1.07$ to $1.51$. In Fig.\ref{mhh.Fig} we depict the LO, total QCD corrected distributions of the final Higgs pair invariant mass $m(h, h)$ and the corresponding $K$-factor. The total QCD correction increases considerably the LO distribution of the Higgs pair invariant mass in the whole plotted range. The peaks for both the LO and total QCD corrected $m(h,h)$ distributions are located in the vicinity of $m(h,h) \sim 400~{\rm GeV}$. The corresponding $K$-factor varies from to $1.17$ to $1.68$.
\begin{figure*}
\begin{center}
\includegraphics[scale=0.6]{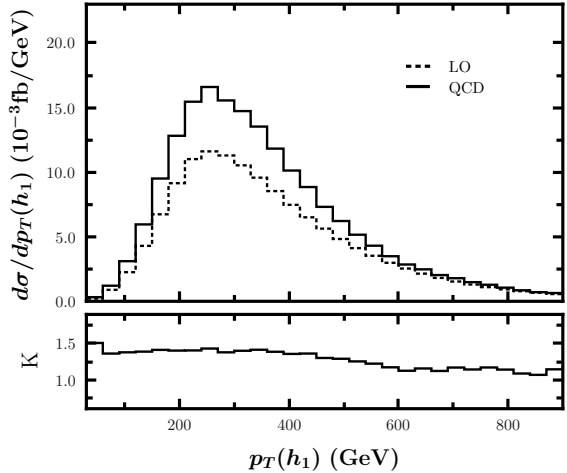}
\caption{The LO, total QCD corrected distributions of the leading Higgs transverse momentum and the corresponding $K$-factor for the $pp \to Z_H Z_H \to A_H  A_H h h+X $ process at the $\sqrt{s}=14~{\rm TeV}$ LHC. }
\label{pth.Fig}
\end{center}
\end{figure*}
\begin{figure*}
\begin{center}
\includegraphics[scale=0.6]{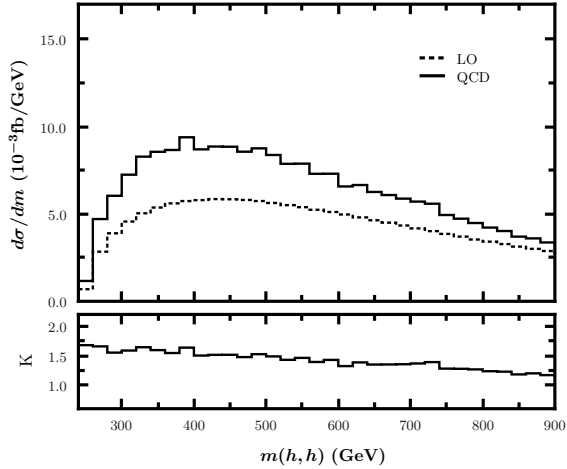}
\caption{The LO, total QCD corrected distributions of the Higgs pair invariant mass and the corresponding $K$-factor for the $pp \to Z_H Z_H \to A_H  A_H h h +X$ process at the $\sqrt{s}=14~{\rm TeV}$ LHC. }
\label{mhh.Fig}
\end{center}
\end{figure*}

\par
{\bf B. Exclusive four-$b$-jet event selection scheme }
\par
We take the event with final four $b(\bar{b})$ jets plus missing energy as the $Z_H$-pair production signal. This $Z_H$-pair production signal process can be denoted as $pp \to Z_H Z_H \to A_H A_H h h \to A_H A_H b\bar{b}b\bar{b}+X$. In the related jet event analysis, we use the Cambridge/Aachen (C/A) jet algorithm \cite{Cacciari:2008gp,jet algorithm} provided in the \texttt{FastJet} package\cite{Cacciari:2011ma} to handle jet combination with the resolution parameter $R=0.4$. In the momentum recombination procedure, the four-momentum of the merged $i$ th and $j$ th jets is obtained by $p_{\mu}^{ij}=p_{\mu}^i+p_{\mu}^j$. After applying the C/A jet algorithm, we are interested in three kinds of prototype events for the signal and SM background: (1) the proto-four-$b$-jet event, (2) the proto-five-$b$-jet event, and (3) the proto-five-jet event with four proto-$b$ jets plus a light-quark proto-$q$ jet $(q=u,\bar{u},d,\bar{d},c,\bar{c},s,\bar{s})$. We apply the following exclusive four-$b$-jet event selection criterion to collect the signature and SM background events:

\par
(1) For the proto-four-$b$-jet events, we accept the event with all the four $b$ jets satisfying \begin{align}\label{jet-constrants-1}
p_T(b)~>~20 ~{\rm GeV}, ~~  |y(b)|~ <~ 2.5, ~~ \Delta R_{bb}~ >~ 0.4.
\end{align}
We call the $b$-jet satisfying conditions of (\ref{jet-constrants-1}) a resolved $b$ jet.

\par
(2) For the proto-five-$b$-jet events, we accept the event with only four resolved $b$ jets and the remained one is rejected by not satisfying one of the (\ref{jet-constrants-1}) constraints.

\par
(3) For the proto-five-jet event with four proto-$b$ jets and a light-quark proto-$q$ jet, we accept the event with four resolved $b$ jets and the light-quark $q$ jet is rejected due to satisfying one of the following limitations:
\begin{align}\label{jet-constrants-2}
p_T(q)~<~20 ~{\rm GeV}, ~~ |y(q)|~ >~ 2.5, ~~ \quad \Delta R_{qb}~ <~ 0.4.
\end{align}
In the expressions of (\ref{jet-constrants-1}) and (\ref{jet-constrants-2}), $p_T(b)$ and $y(b)$ ($p_T(q)$ and $y(q)$) are the transverse momentum and rapidity of the $b$ jet (light-quark $q$ jet), respectively, and $\Delta R_{bb}$ ($\Delta R_{qb}$) is the separation in the plane of azimuthal angle and rapidity between two $b$ jets (between the light-quark $q$ jet and a $b$ jet).

\par
After applying the above exclusive four-$b$-jet event selection criterion, we can obtain the signal event including four $b$ jets plus missing energy, and the background event with only four $b$ jets. We call the $b$ jet with the largest transverse momentum among the four $b$ jets as the leading $b$ jet, denoted as $b_1$, and that with the second largest transverse momentum $b$ jet as the subleading $b$ jet, denoted as $b_2$. In a similar way we name the Higgs boson with the largest transverse momentum among the two Higgs bosons as the leading Higgs boson, denoted as $h_1$, and the remaining Higgs boson is called the subleading Higgs boson, denoted as $h_2$. Namely, we have $p_T(b_1) > p_T(b_2) > p_T(b_{3,4})$ and $p_T(h_1) > p_T(h_2)$.

\par
{\bf C. Differential cross sections for $pp\to Z_H Z_H \to A_H A_H b \bar{b} b \bar{b}+X$ }
\par
In the following event analysis related to final jets we adopt the above exclusive four-$b$-jet event selection criterion. We present the LO, total QCD corrected distributions of the transverse momentum and rapidity of the leading $b$ jet and the corresponding $K$-factors for the accepted signal events of $pp \to Z_H Z_H \to A_H  A_H h h \to A_H A_H b \bar{b} b \bar{b}+X$ process at the $\sqrt{s}=14~{\rm TeV}$ LHC in Fig.\ref{ptb1.Fig} and Fig.\ref{yb1.Fig}, separately. We see from Fig.\ref{ptb1.Fig} that the total QCD correction significantly enhances the LO distribution in the low $p_T(b_1)$ range, and the LO and total QCD corrected distributions have similar line shapes as each other. Both the LO and total QCD corrected distributions reach their maximal values at the position of $p_T(b_1)\sim 200~{\rm GeV}$ with $K = 1.22$, and the $K$-factor is in the range between $1.00$ and $1.24$. From Fig.\ref{yb1.Fig} we can see the total QCD corrected distribution has obvious enrichment to the LO distribution, and the $K$-factor varies from $1.08$ to $1.15$  in the plotted rapidity region.

\par
We depict LO, total QCD corrected differential cross sections of $m(b_1,b_2)$, which represents the invariant mass of the leading and subleading $b$ jets, and the corresponding $K$-factor for the accepted signal events of the $pp \to Z_H Z_H \to A_H  A_H h h \to A_H A_H b \bar{b} b \bar{b}+X$ process at the $\sqrt{s}=14~{\rm TeV}$ LHC in Fig.\ref{mbb.Fig}. From the figure we see that the total QCD corrected distribution resembles that at the LO, and both the LO and total QCD corrected distributions reach their maximal values at the position of $m(b_1,b_2)\sim 300~{\rm GeV}$ with $K=1.22$. It shows that the total QCD correction enhances the LO distribution, and the $K$-factor varies from $1.00$ to $1.25$ in the plotted invariant mass range.
\begin{figure*}
\begin{center}
\includegraphics[scale=0.6]{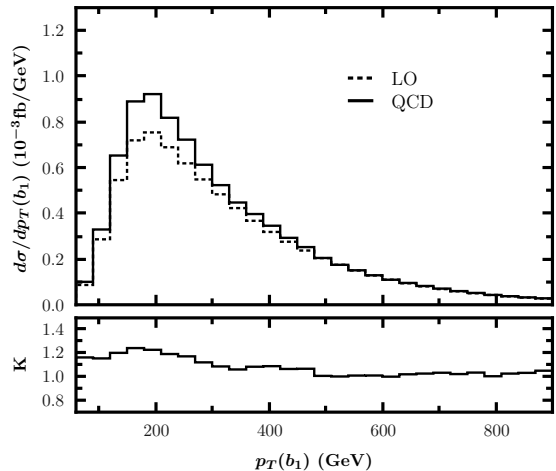}
\caption{The LO, total QCD corrected distributions of the transverse momentum of the leading $b$ jet ($b_1$) and the corresponding $K$-factor for the $pp \to Z_H Z_H \to A_H  A_H h h \to A_H A_H b \bar{b} b \bar{b}+X$ process at the $\sqrt{s}=14~{\rm TeV}$ LHC by adopting the exclusive four-$b$-jet event selection scheme.}
\label{ptb1.Fig}
\end{center}
\end{figure*}
\begin{figure*}
\begin{center}
\includegraphics[scale=0.6]{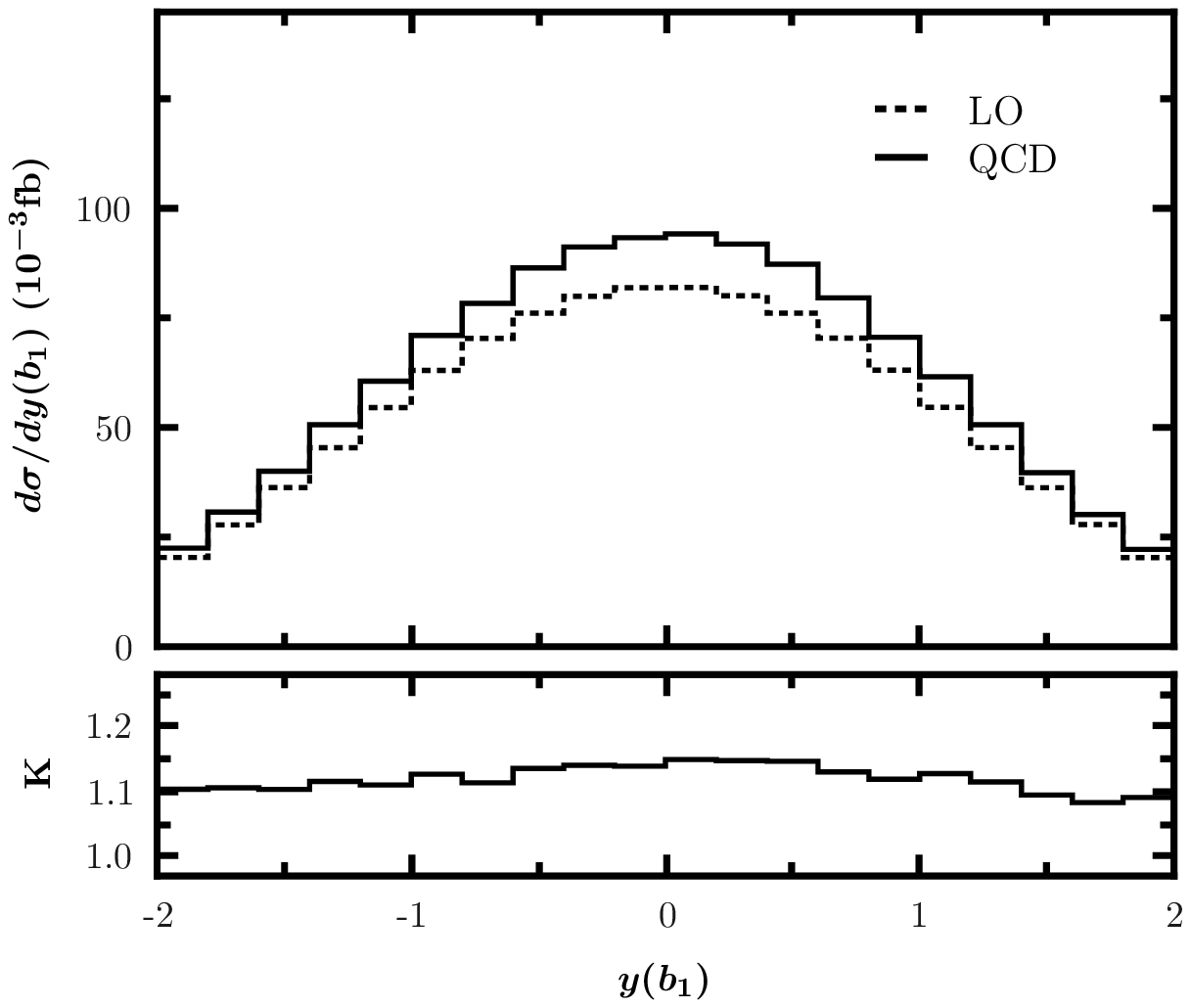}
\caption{The LO, total QCD corrected distributions of rapidity of the leading $b$ jet ($b_1$) and the corresponding $K$-factor for the $pp \to Z_H Z_H \to A_H  A_H h h \to A_H A_H b \bar{b} b \bar{b}+X$ process at the $\sqrt{s}=14~{\rm TeV}$ LHC by adopting the exclusive four-$b$-jet event selection scheme.}
\label{yb1.Fig}
\end{center}
\end{figure*}
\begin{figure*}
\begin{center}
\includegraphics[scale=0.6]{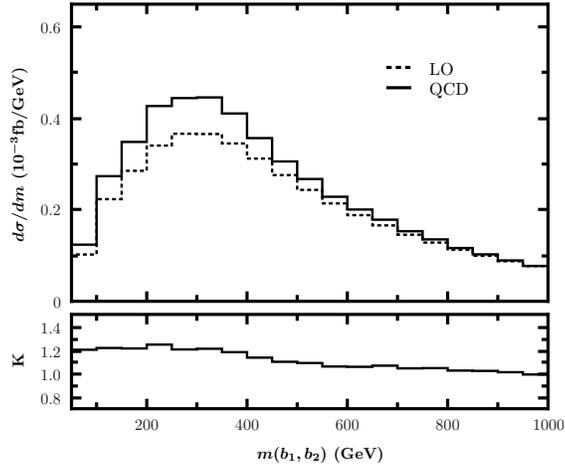}
\caption{The LO, total QCD corrected distributions of the invariant mass of the leading and subleading $b$ jets and the corresponding $K$-factor for the $pp \to Z_H Z_H \to A_H  A_H h h \to A_H A_H b \bar{b} b \bar{b}+X$ process at the $\sqrt{s}=14~{\rm TeV}$ LHC by adopting the exclusive four-$b$-jet event selection scheme. }
\label{mbb.Fig}
\end{center}
\end{figure*}

\par
Corresponding to the signature of the $Z_H$-pair production at the LHC, the possible main SM backgrounds are from the $pp \to b \bar{b} b \bar{b}+X$ and $pp \to Zb \bar{b} \to b\bar{b}b\bar{b}+X$ processes with four resolved $b$ jets. Refs.{\cite{Greiner,Binoth}} show that the NLO QCD corrections to the $pp \to b \bar{b} b \bar{b}+X$ process at the LHC lead to an enhancement of the integrated cross section for the $pp \to b \bar{b} b \bar{b}+X$ process at the central scale by roughly $50\%$, and considerably improve the prediction. That means if we include the NLO QCD corrections in the $pp \to b \bar{b} b \bar{b}+X$ background process, the ratio of signal to background will become smaller. We define parameter $H_T$ as $H_T = \sum_{i} |\vec{p}_T(i)|$, which is the scalar sum of the transverse momenta of all the final four $b$ jets (and $\slashed{E}_T$) for the background (signal) event. In Fig.\ref{Ht.Fig} we present the normalized LO, total QCD corrected $H_T$ distributions for the signal by adopting the exclusive four-$b$-jet selection scheme mentioned above, the NLO QCD corrected distribution for the $pp \to b \bar{b} b \bar{b}+X$ process (the data are taken from Fig.3 and Table I in Ref.\cite{Worek} with the Nagy-Soper subtraction formalism for real radiation at NLO and the selection criterion declared in that paper), and LO distribution for the $pp \to Z b \bar{b} \to b \bar{b} b \bar{b}+X$ background by using the same four-$b$-jet selection scheme as used for the signal process. There the distributions are normalized by the corresponding LO total cross sections. We can see from Fig.\ref{Ht.Fig} that the LO $H_T$ distribution for the signal process is obviously increased by the total QCD correction. That means when we adopt the exclusive four-$b$-jet event selection criterion mentioned above, the total QCD correction enhances the LO $H_T$ differential cross section in the plotted $H_T$ range. We can see also that the SM background events tend to be concentrated in the low $H_T$ region with a peak in the vicinity of $H_T \sim 170~{\rm GeV}$ and then its event number declines quickly, while the LO and total QCD corrected $H_T$ distributions for the $Z_H$-pair production signature have a flatter peak in the vicinity of $H_T \sim 600~{\rm GeV}$ and descend slowly as illustrated in Fig.\ref{Ht.Fig}. That indicates if we take proper lower limits on $H_T$ parameter, the background from both the $pp \to b \bar{b} b \bar{b}+X$ and $pp \to Zb \bar{b} \to b\bar{b}b\bar{b}+X$ processes can be significantly suppressed.
\begin{figure*}
\begin{center}
\includegraphics[scale=0.6]{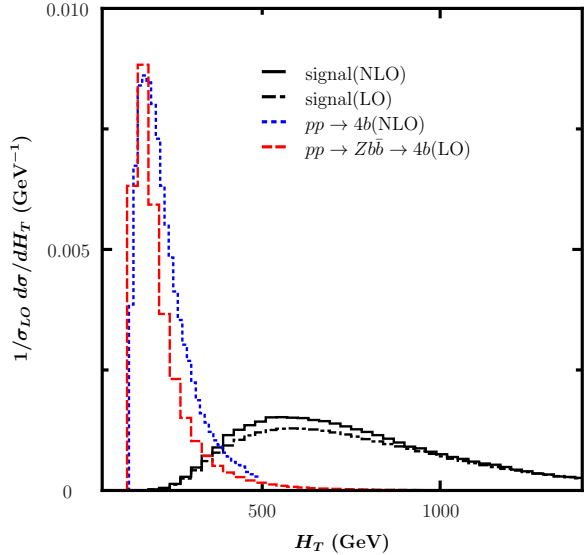}
\caption{The normalized LO, total QCD corrected $H_T$ distributions for the signal process $pp \to Z_H Z_H \to A_H  A_H h h \to A_H A_H b \bar{b} b \bar{b}+X$, the normalized LO $H_T$ distribution for the $pp \to Zb \bar{b} \to b \bar{b} b \bar{b}+X$, and the normalized NLO QCD corrected $H_T$ distribution for the $pp \to b \bar{b} b \bar{b}+X$ at the $\sqrt{s}=14~{\rm TeV}$ LHC. There the normalized NLO QCD corrected $H_T$ distribution for the SM background processes $pp \to b \bar{b} b \bar{b}+X$ is obtained from Ref.\cite{Worek}.  }
\label{Ht.Fig}
\end{center}
\end{figure*}

\par
There exists a possibility that the $pp \to Zbj+X \to b \bar{b} b j+X$ and $pp \to Zjj+X \to b \bar{b} jj+X$ productions are the SM background sources where $j$ denotes the gluon/light-quark jet, when a $j$ jet is mistagged as a $b$ jet. Since this mistagging possibility is rather small in experiment, these two channels will slightly enhance the $H_T$ distribution of the background for the $pp \to Zb\bar{b} \to b \bar{b} b \bar{b}+X$ process. From Fig.\ref{Ht.Fig}, we can see that even if we consider the background enhancement due to jet mistagging, it is still possible to efficiently suppress the background by putting proper lower constraints on $H_T$.

\par
\section{SUMMARY}
\par
In this paper, we present the calculations of the $Z_H$-pair production at the $\sqrt{s}=14~{\rm TeV}$ LHC including the pure NLO QCD correction and the $gg$-fusion contribution within the framework of the littlest Higgs model with $T$ parity. By implementing the diagram subtraction scheme, we separate out the $Z_H Z_H$ and $Z_H q_-$ production channels and recover convergence of the perturbative description for the $Z_H$-pair production process. The renormalization/factorization scale dependence of the integrated cross section is investigated, and we find that the total QCD correction slightly reduces the scale uncertainty. We present some kinematic distributions of the final products considering the subsequential on shell $Z_H$ decays of $Z_H \to A_H h \to  A_H b\bar{b}$ that provides an interesting channel for the di-Higgs production. The analyses for the kinematic distributions of final particles show that with our exclusive four-$b$-jet event selections scheme, the total QCD correction considerably increases the LO distributions. We compare the $H_T$ distributions for the $Z_H$-pair production signal and the SM background, and conclude that they are remarkably different. The di-Higgs boson signal events via the $Z_H$-pair production can be discriminated from the possible SM background by taking proper cuts on the $H_T$ parameter.

\vskip 5mm
\par
\noindent{\large\bf Acknowledgments:} This work was supported in part by the National Natural Science Foundation of China (Grants. No.11275190, No.11375008, No.11375171, and No.11347101) and the Startup Foundation for Doctors of Kunming University of Science and Technology under Grant No.KKSY201356060.

\end{document}